\documentclass[11pt]{article}

\usepackage{amsmath}
\usepackage{amssymb}
\usepackage{amsthm}
\usepackage{latexsym}
\usepackage{color}
\usepackage{float}
\usepackage{graphicx}
\usepackage{color}
\usepackage{epstopdf}
\usepackage{enumerate}
\usepackage[T1]{fontenc}
\usepackage[english]{babel}
\usepackage[table]{xcolor}

\DeclareSymbolFont{calletters}{OMS}{cmsy}{m}{n}
\DeclareSymbolFontAlphabet{\mathcal}{calletters}

\def\be{\begin{eqnarray}}
\def\ee{\end{eqnarray}}

\def\b*{\begin{eqnarray*}}
\def\e*{\end{eqnarray*}}

\newtheorem{Theorem}{Theorem}[part]

\newtheorem{Proposition}{Proposition}[part]

\newtheorem{Lemma}{Lemma}[part]

\makeatletter \@addtoreset{equation}{section}

\@addtoreset{Definition}{section}

\@addtoreset{Theorem}{section}

\@addtoreset{Proposition}{section}

\@addtoreset{Property}{section}

\@addtoreset{Assumption}{section}

\@addtoreset{Corollary}{section}

\@addtoreset{Lemma}{section}

\@addtoreset{Remark}{section}

\@addtoreset{Example}{section}

\@addtoreset{Condition}{section}

\def \N{\mathbb{N}}

\def \R{\mathbb{R}}

\def\={\;=\;}
\def\.{\;.}

\def\1{{\bf 1}}

\def\b*{\begin{eqnarray*}}
\def\e*{\end{eqnarray*}}

 \def\normeL2#1{\left\|{#1}\right\|_{L^2}}

\newtheorem{rem}{Remark}

\begin{document}
\title{A convex duality method for optimal liquidation with participation constraints\footnote{This research has been conducted with the support of the Research Initiative ``Mod\'elisation des march\'es financiers \`a haute fr\'equence'' (formerly ``Ex\'ecution optimale et statistiques de la liquidit\'e haute fr\'equence'') under the aegis of the Europlace Institute of Finance. This paper has been improved following the reports of two anonymous referees who really need to be thanked.}}

\author{Olivier Gu\'eant\footnote{Universit\'e Paris-Diderot, UFR de Math\'ematiques, Laboratoire Jacques-Louis Lions. Avenue de France, 75013 Paris.}, Jean-Michel Lasry\footnote{Universit\'e Paris-Dauphine, Ceremade. Place du Mar\'echal de Lattre de Tassigny, 75116 Paris.}, Jiang Pu\footnote{Institut Europlace de Finance. 28 Place de la Bourse, 75002 Paris.}}

\date{}

\maketitle

\begin{center}
\textbf{Abstract}
\end{center}

In spite of the growing consideration for optimal execution in the financial mathematics literature, numerical approximations of optimal trading curves are almost never discussed. In this article, we present a numerical method to approximate the optimal strategy of a trader willing to unwind a large portfolio. The method we propose is very general as it can be applied to multi-asset portfolios with any form of execution costs, including a bid-ask spread component, even when participation constraints are imposed. Our method, based on convex duality, only requires Hamiltonian functions to have $C^{1,1}$ regularity while classical methods require additional regularity and cannot be applied to all cases found in practice.

\section{Introduction}

When he is willing to unwind a large portfolio, a trader faces a trade-off between market risk on the one hand and market impact and execution costs on the other hand. Selling rapidly a large quantity of shares is indeed costly as it requires to take liquidity from limit order books. Selling slowly however exposes to price risk because of other market participants' actions. If a trader makes the decision to unwind his portfolio at a given time, his decision is certainly based on market prices at that time. He needs therefore to sell fast enough so that prices do not move too much over the course of the execution process.\\

Optimal execution has been an important topic in the academic literature for around 15 years. The above trade-off between execution costs and market impact on the one hand and price risk on the other hand has indeed been modeled in the seminal papers \cite{almgren1999value,almgren2001optimal} of Almgren and Chriss published in 1999 and 2001. Almgren and Chriss proposed a simple framework to solve the problem of optimally scheduling the execution process and this model has largely been used and extended since then, by both practitioners and academics.  Initially written in the one-asset case, in discrete time, with quadratic execution costs, and with a Bachelier dynamics for the price, the Almgren-Chriss model has then been considered in continuous time and generalized to allow for more realistic execution costs and random execution costs -- see \cite{almgren2003optimal}. Black-Scholes dynamics for the price has also been considered -- see \cite{gatheral2011optimal} -- and attempts to generalize the model in other directions have been made in the one-asset case, for instance to take account of stochastic volatility and stochastic liquidity, see \cite{almgren2011optimal}. Discussions on the optimization criteria and their consequences on optimal strategies have also an important place in the literature, see for instance  \cite{almgren2007adaptive, forsyth2009optimal, jk, lorenz2010mean, tse2011comparison}, and the very interesting results in the case of increasing or decreasing risk aversion obtained in \cite{schied2009risk}. Important extensions to multi-asset portfolios have been developed for instance by Schied and Sch\"oneborn \cite{schied2010optimal}. Gu\'eant also developed a general framework for optimal liquidation in \cite{gueantframework} and used it to price block trades.\\

In this classical literature on optimal execution, market microstructure is modeled through permanent market impact and execution cost (or instantaneous market impact) functions. These functions account for the influence of the trader on the stock price and for the costs paid by the trader to take liquidity from the market. Another strand of literature has developed following Obizhaeva and Wang, whose paper \cite{obizhaeva2005optimal} has been a preprint since 2005. In this part of the literature, the very dynamics of the order book is modeled in order to introduce transient market impact into optimal execution problems. However, in the post-MIFiD and post-Reg NMS world, with access to several multilateral trading facilities and to various dark pools, a relevant approach of this kind would require to model all venues... and this would lead to a rather complicated model! Models \emph{\`a la} Almgren-Chriss are simpler and permit to sum up the influence of the different venues (associated with different microstructures) into one or two functions.\\

In most of the papers in the literature, the optimal strategy does not depend on the evolution of prices (see in particular \cite{schied2010optimal}). The outcome of models \emph{\`a la} Almgren-Chriss is indeed an optimal trading curve, stating, before it starts, the optimal time scheduling of the execution process. This trading curve constitutes the first (strategic) layer of most execution algorithms. An execution algorithm is indeed usually made of two layers:\footnote{See \cite{ll} for a general description of execution algorithms. See also \cite{bouchard2009optimal} for another viewpoint.} a strategic layer, which controls the risk with respect to a benchmark and mitigates execution costs; and a tactical layer, which seeks liquidity inside order books, through all types of orders, and across all other liquidity pools (lit or dark). The second layer has been studied in the literature more recently through new models involving limit orders (see for instance \cite{bayraktar2012liquidation,gueant2012general,gueant2012optimal,laruellelimit}) or through the study of liquidation with dark pools (see \cite{kratz2009optimal,kratz2012optimal} and \cite{laruelledark}).\\

In this article, we focus on the first layer of execution algorithms for a multi-asset portfolio.\footnote{The case of a multi-asset portfolio is also interesting for one-asset portfolio liquidation as it enables to consider a round trip on an additional asset for hedging.} We consider a Von Neumann-Morgenstern expected utility framework in the case of an investor with constant absolute risk aversion. We consider a general form of execution costs and the optimal strategy is characterized by a Hamiltonian system as in \cite{gueantframework} and \cite{schied2010optimal}. Numerical approximations of the optimal strategy are briefly discussed in \cite{gueantframework} but the method presented in \cite{gueantframework} is limited to a small class of Hamiltonian functions (strictly convex functions with $C^2$ regularity). The method we present in this paper is more general as it only requires Hamiltonian functions to have $C^{1,1}$ regularity. It enables to solve numerically the problem of the optimal strategy to unwind a multi-asset portfolio when bid-ask spreads are taken into account and when one adds participation rate constraints (an upper bound to the volume that can be traded, relative to market volume).\\

Numerical methods are very rarely discussed in the literature (one important exception is \cite{lale}). In fact, the problem is rather simple in the one-asset case as the bid-ask spread component plays no role and as most papers do not consider participation constraints. In the one-asset case, a shooting method can indeed be used to obtain a precise numerical approximation of the optimal trading curve. In the multi-asset case, however, shooting methods are not relevant anymore. Newton's methods to find a solution to the Hamiltonian equations or to the Euler-Lagrange equations associated to Almgren-Chriss-like models are not possible either, except in the case of smooth functions... too smooth to embed a bid-ask spread component or participation constraints. Considering the Bellman equations associated to the optimal control is always possible, at least theoretically, but it has serious drawbacks in terms of computation time and numerical precision. The method we present is based on convex duality and allows to consider all conceivable practical cases as it only requires Hamiltonian functions to be $C^{1,1}$.\\

We present the general framework in continuous time in Section 2 and we state the classical existence and uniqueness results for the optimal liquidation strategy. We also provide the Hamiltonian equations that characterize this optimal liquidation strategy. These equations, and their discrete counterparts, play a major role in our paper. In Section 3, we present our numerical method (based on convex duality) to approximate the optimal liquidation strategy, and we prove a convergence result. In Section 4, we present numerical examples. Proofs are presented in Appendix A. Appendix B is dedicated to the discrete counterpart of the continuous-time model presented in Section 2.

\section{Optimal liquidation: setup and classical results}
\subsection{Setup}

We consider a portfolio of $d$ different stocks with initial quantities $q_0 = (q_0^1, \ldots, q_0^d)$. We consider the problem of unwinding this portfolio over the time window\footnote{We consider here a variant of the classical models in continuous time used in \cite{gueantframework} or \cite{schied2010optimal}. The discrete counterpart of our model is presented in Appendix B.} $[0,T]$ (the time horizon being usually a few minutes to a few hours).\\

We consider a probability space $(\Omega, \mathbb{P})$ equipped with a filtration $(\mathcal{F}_t)_{t\in [0,T]}$ satisfying the usual conditions. We assume that all stochastic processes are defined on $(\Omega, (\mathcal{F}_t)_{t\in [0,T]}, \mathbb{P})$. We also introduce the set $\mathcal{P}(0,T)$ of progressively measurable processes defined on $[0,T]$.\\

Market volume processes are denoted by $(V^1_t,\ldots,V^d_t)_{t \in [0,T]}$. They are assumed to be deterministic ($\mathcal{F}_0-$measurable), positive and bounded. For each stock, we consider a maximum participation rate and we denote them by  $\rho^1_{m},\ldots,\rho^d_m \in \mathbb{R}_+^*$. This allows to define the set of admissible liquidation strategies:
$$\mathcal{A} = \Bigg\lbrace (v^1_t, \ldots, v^d_t)_{t\in [0,T]} \in \mathcal{P}(0,T), \forall i, |v^i_t| \le \rho^i_m V ^i_t \textrm{\;a.e.\; on } [0,T]\times\Omega,\forall i, \int_0^T v^i_t dt = q_0^i \textrm{\;a.s.}  \Bigg\rbrace.$$ For a liquidation strategy $(v^1_t, \ldots, v^d_t)_{t\in [0,T]} \in \mathcal{A}$, representing the velocity at which the trader sells his shares, we denote by $(q^1_t, \ldots, q^d_t)_{t \in [0,T]}$ the process giving the state of the portfolio. It verifies:

$$\forall i, q^i_t = q^i_0 - \int_0^t v^i_s ds.$$

\begin{rem}
In other words, our hypotheses on the strategies $(v^1, \ldots, v^d)$ are simply that one cannot trade too quickly, relatively to market volume, and that we indeed liquidate the portfolio by time $T$.
\end{rem}
\begin{rem}
We always assume that liquidation is feasible in the sense that $$\forall i, |q^i_0| \le  \rho^i_m \int_0^T V^i_t dt.$$
\end{rem}

For each stock, we consider Brownian dynamics for the price:\footnote{We do not consider permanent market impact as it plays no role in liquidation strategies -- see \cite{gatheral2010no} and \cite{gperm}.}

$$\forall i, dS^i_t = \sigma^i dW^i_t \qquad \sigma^i >0,$$ and we assume that the $d$-dimensional Brownian motion $(\sigma^1 W^1, \ldots, \sigma^d W^d)$ has a covariance matrix $\Sigma$ that is not singular.\\

\begin{rem}
Considering Brownian dynamics for prices instead of Black-Scholes dynamics has almost no influence in terms of modelling, as we consider time horizons of at most a few hours. From a mathematical point of view however, the former is more practical than the latter.
\end{rem}

\begin{rem}
The case of non-constant volatility processes can easily be considered in our model, as soon as these processes remain deterministic and positive. For instance one may relate the volatility processes to the market volume processes.
\end{rem}

The price obtained by the trader at time $t$ for his trades on stock $i$ is not $S^i_t$ because of execution costs. To model these execution costs, we introduce $d$ functions $L^1, \ldots, L^d$ verifying the following hypotheses:\\

\begin{itemize}
\item $ \forall i, L^i(0)=0 $,
\item $ \forall i,  L^i$ is an even function,\footnote{This assumption can easily be relaxed, to include for instance an asymmetrical stamp duty.}
\item $ \forall i, L^i$ is increasing on $\mathbb{R}_+$,
\item $ \forall i, L^i$ is strictly convex.
\end{itemize}

Now, for $(v^1, \ldots, v^d) \in \mathcal{A}$, we define the cash process $(X_t)_{t \in [0,T]}$ by:

$$X_t = \int_0^t \sum_{i=1}^d \left( v^i_s S^i_s - V^i_s L^i\left(\frac{v^i_s}{V^i_s} \right) \right) ds.$$

\begin{rem}
In practice, we need to cover the cases $L^i(\rho) = \eta^i |\rho|^{1+\phi^i} + \psi^i |\rho|$ for $\eta^i >0, \psi^i \ge 0$ and $\phi^i \in (0,1]$. The proportional part of the function corresponds to bid-ask spread, stamp duty and/or financial transaction tax -- we generally call it the bid-ask spread component. The superlinear part is the classical execution cost component of all models \emph{\`a la} Almgren-Chriss.
\end{rem}

Our objective function for $(v^1, \ldots, v^d) \in \mathcal{A}$ is

$$J(v^1, \ldots, v^d) = \mathbb{E}\left[-\exp(-\gamma X_T)\right],$$ where $\gamma>0$ is the absolute risk aversion parameter of the trader.\\

In other words, our problem is to find:

$$(v^{1*}, \ldots, v^{d*}) \in \textrm{argmax}_{(v^1, \ldots, v^d) \in \mathcal{A}}J(v^1, \ldots, v^d).$$

Adapting the results obtained in  \cite{gueantframework} and \cite{schied2010optimal}, we can show that there exists such an optimal liquidation strategy. We can also show that stochastic strategies cannot do better than deterministic ones. These results, along with a characterization of the unique deterministic optimal strategy, are exhibited in the next paragraphs.

\subsection{Optimal liquidation strategy}

The results obtained in \cite{gueantframework} and \cite{schied2010optimal} can easily be adapted to the case considered here. First of all, one can restrict the set of  liquidation strategies to deterministic processes in $\mathcal{A}$. We denote by $\mathcal{A}_\mathrm{det}$ this set, which consists of the liquidation strategies in $\cal A$ that are $\mathcal{F}_0$-measurable.\\

Our first Proposition states that, in the case of deterministic strategies, the objective function simplifies, since final wealth $X_T$ is normally distributed.

\begin{Proposition}
\label{gaussian}
If $(v^1, \ldots, v^d) \in \mathcal{A}_{\mathrm{det}}$ then $X_T$ is normally distributed, and
$$J(v^1, \ldots, v^d) = -\exp\left(-\gamma\left(q_0' S_0 - \sum_{i=1}^d \int_0^T V^i_s L^i\left(\frac{v^i_s}{V^i_s}\right) ds -  \frac 12 \gamma \int_0^T  q_s' \Sigma q_s ds \right)\right).$$
\end{Proposition}

A consequence of this Proposition is that the problem boils down to solving the following variational problem:

$$(\mathcal P) \quad \inf_{(q^1,\ldots, q^d) \in \mathcal{C}}I(q^1,\ldots, q^d)$$ where $$I(q^1,\ldots, q^d) = \sum_{i=1}^d \int_0^T V^i_s L^i\left(\frac{\dot{q}^i(s)}{V^i_s}\right) ds +  \frac 12 \gamma \int_0^T  q(s)' \Sigma q(s) ds,$$ and where $$\mathcal{C} = \left\lbrace q \in W^{1,1}((0,T),\mathbb{R}^d), q(0) = q_0, q(T) = 0, \forall i, |\dot{q}^i(t)| \le \rho^i_m V^i_t, \textrm{\;a.e.\;in\;} (0,T)\right\rbrace.$$

Existence and uniqueness of minimizers for $I$ are obtained using classical techniques of variational calculus and convex optimization as in \cite{gueantframework}. Moreover, the optimal liquidation strategy is characterized by a Hamiltonian system.

\begin{Theorem}[Existence, uniqueness and characterization of an optimal strategy]
\label{main}
There exists a unique function $q^* \in \mathcal{C}$ that minimizes the function $I$ defined in problem $(\mathcal P)$.\\
There exists a $W^{1,1}$ function $p$ such that $(q^*,p)$ is a solution of the Hamiltonian system
\[
(S_H):\qquad \left \{
\begin{array}{c c l}
    \dot{p}(t) & = & \gamma \Sigma q(t) \\
    \dot{q}^i(t) & = & V^i_t {H^i}'(p^i(t)), \quad \forall i \in \{ 1, \ldots, d\}  \\
\end{array}
\right.
\]
with boundary conditions $q(0) = q_0, \quad q(T) = 0$, where:
$$H^i(p) = \sup_{|\rho| \le \rho^i_m}  p \rho -   L^i\left(\rho\right),$$

Moreover, if a pair $(q,p)$ of $W^{1,1}$ functions is solution of $(S_H)$, then $q=q^*$.
\end{Theorem}

\begin{rem}
Although there is uniqueness of the optimal trajectory $q^*$, there may not be uniqueness of $p$. This turns out to be important when it comes to numerics.
\end{rem}

\begin{rem}
If market volume processes are continuous, it is clear that for any solution $(p,q)$ of the system $(S_H)$, $q$ has $C^1$ regularity and $p$ has $C^2$ regularity. We shall therefore approximate $p$ instead of $q$.
\end{rem}

The Hamiltonian characterization $(S_H)$ is more suited to numerics than the Euler-Lagrange equation used in most papers since the execution cost functions $L^i$s are not of class $C^1$ as soon as there is a bid-ask spread component. Finding an approximation to a solution of the Hamiltonian system $(S_H)$ (or in fact to the counterpart of this system in a discrete-time model) is the goal of this paper and we present our method in the next section. But before turning to our numerical method, let us conclude this section by stating that stochastic strategies cannot do better than the best deterministic strategy.

\begin{Theorem}[Optimality of deterministic strategies]
\label{optisdet}
$$\sup_{(v^1, \ldots, v^d) \in \mathcal{A}} \mathbb{E}\left[-\exp\left(-\gamma X_T\right)\right] =\sup_{(v^1, \ldots, v^d) \in \mathcal{A}_\mathrm{det}} \mathbb{E}\left[-\exp\left(-\gamma X_T\right)\right].$$
\end{Theorem}

\section{Numerical method: convex duality to the rescue}

\subsection{Preliminary remarks}

The model we have presented in the above section is a model in continuous time. Continuous-time models are useful to benefit from differential calculus and to have intuition on numerical methods. However, as explained in the introduction, the problem we tackle consists of the first layer of execution algorithms and a decision has to be taken every 5 or 10 minutes. Hence, the problem faced by practitioners is naturally in discrete time rather than in continuous time.
The goal of this section is to present a new numerical method for approximating a solution $(p,q)$ of the Hamiltonian system $(S_H)$, or more exactly of a discrete version of this system. This discrete version $(\tilde{S}_H)$ corresponds to the optimality conditions of the discrete counterpart of the optimization problem of Section 2 (see Appendix B for the presentation of the discrete-time model).
Considering a time grid $0, \Delta t, \ldots, N\Delta t = T$, we are looking for $((q^i_n)_{1\le i \le d, 0 \le n \le N},(p^i_n)_{1\le i \le d, 0 \le n \le N-1})$ satisfying:

\[
(\tilde{S}_H): \quad \left \{
\begin{array}{c c l}
    p_{n+1} & = & p_n + \Delta t \gamma \Sigma q_{n+1}, \quad 0\le n <N-1\\
    q^i_{n+1} & = & q^i_n  + \Delta t V^i_{n+1} {H^i}'(p^i_n) , \quad 0\le n <N, \quad \forall i \in \{ 1, \ldots, d\}\\

\end{array}
\right.\] $$ q_0 = q_0, \quad q_N = 0.$$

The first part of the system is made of linear equations. The difficulty then relates to the nonlinearity introduced by the Hamiltonian functions $H^i$s. In particular, it is noteworthy that in the initial Almgren-Chriss case \cite{almgren1999value,almgren2001optimal} with a quadratic execution cost function, the system boiled down to a linear one.\\

In practice, a classical form for the execution cost functions $L^i$s is  $$L(\rho) = \eta |\rho|^{1+\phi} + \psi |\rho|, \qquad  \eta >0, \psi\ge 0, \phi \in (0,1].$$
The superlinear component has been considered since the initial papers by Almgren and Chriss. In \cite{almgren1999value,almgren2001optimal}, the authors considered the quadratic case $\phi=1$ to obtain closed form solutions. However, realistic values for $\phi$ are usually estimated to be between 0.4 and 0.8. For instance, Almgren and coauthors from Citigroup estimated $\phi \simeq 0.6$ using a large dataset of meta-orders (see \cite{almgrenTC}). The proportional component models the influence of the bid-ask spread, of a stamp duty, or of a financial transaction tax.\\

The Hamiltonian function $H$ associated to $L(\rho) = \eta |\rho|^{1+\phi} + \psi |\rho|$ with participation constraint $\rho_m$ is:

$$H(p) = \sup_{|\rho| \le \rho_m}  p \rho -   \eta |\rho|^{1+\phi} - \psi |\rho|  = \begin{cases} 0 &\mbox{if } |p| \le \psi \\
\phi\eta\left(\frac{|p|-\psi}{\eta(1+\phi)}\right)^{1+\frac 1\phi}& \mbox{if }  \psi <|p| \le \psi + \eta(1+\phi)\rho_m^\phi\\
(|p|-\psi)\rho_m - \eta \rho_m^{1+\phi}                  & \mbox{if }  \psi + \eta(1+\phi)\rho_m^\phi < |p|
\end{cases}$$

As $L$ is strictly convex, $H$ is $C^1$ with:

$${H}'(p) = \textrm{sign}(p)  \min\left(\rho_{m}, \left(\frac{\max(|p|-\psi,0)}{\eta(1+\phi)}\right)^{\frac{1}{\phi}}\right)$$

\begin{figure}[H]
\centering
\includegraphics[width = 0.7\textwidth]{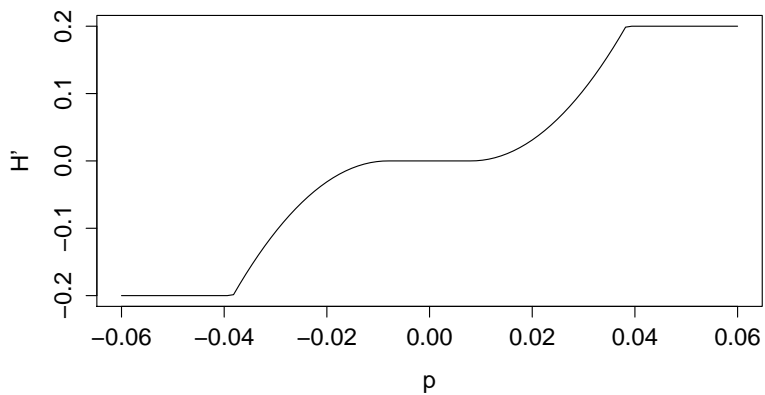}
\caption{Shape of $H'$ for $\eta=0.045$, $\phi=0.5$, $\psi=0.0081$ and $\rho_m=20\%$.}
\label{jiang}
\end{figure}

This function $H'$ is not differentiable (see Figure \ref{jiang}). This is an issue to build a general numerical method since the systems $(S_H)$ and $(\tilde{S}_H)$ involve derivatives of Hamiltonian functions of this form.\\

In dimension $d=1$, solutions of the system $(\tilde{S}_H)$ are easy to approximate numerically using simple shooting methods. One can indeed replace the terminal condition $q_N=0$ by an initial condition on $p$ of the form $p_0=\lambda$. Then, the system can easily be solved recursively (forward in time). It remains to notice that $q_N$ is a monotone function of $\lambda$ to end up with an efficient method that finds the appropriate value of $\lambda$ such that $q_N=0$. However, as soon as we consider several assets, the issue of numerical methods arises. The shooting method described above is not relevant anymore: the monotonicity property is lost and usual methods to find the appropriate value of $p_0$ (such as gradient descents) generally fail.\\

Using a Newton's method on the whole system, as discussed in \cite{gueantframework}, is only possible when the functions $H^i$s are twice differentiable. Therefore, it cannot be used to find an approximate solution of the system $(\tilde{S}_H)$ when there is a bid-ask spread component or when participation constraints are considered.\\

Another method would be to consider the Bellman equations associated to the problem and to use classical techniques to approximate the solutions of these equations. This is always possible, independently of the regularity of the functions $L^i$s and $H^i$s. However, this method has several serious drawbacks. First, Bellman equations need to be solved on a $d$-dimensional grid because dynamic programming requires to be able to solve execution problems for (almost) all values of the state variable $q$. In terms of computation time, this is not an efficient method, especially as $d$ becomes larger than $2$ or $3$. Moreover, the domain for the state variable $q$ need to be considered large, as optimal strategies may involve round trips for hedging purposes. Even in dimension $1$ or $2$, solving the Bellman equations associated to $(\tilde{S}_H)$ is not recommended as optimal strategies must either be on the grid, or approximated using interpolation methods (\emph{e.g.} splines)... and the step of the grid in $q$ has no reason to be the same over the course of the execution process if one accounts for the variability of the market volume process.\\

The method we propose is based on a dual formulation of problem $(\mathcal{P})$. It only requires the derivative of the functions $H^{i}$s to be Lipschitz (and this is the case for the form considered above). In addition, our method is iterative and only requires computations on a vector of size $N\times d$.\\

\subsection{Dual problems}

The problem $(\mathcal P)$ we considered in Section 2 is:

$$\inf_{(q^1,\ldots, q^d) \in \mathcal{C}} \sum_{i=1}^d \int_0^T V^i_s L^i\left(\frac{\dot{q}^i(s)}{V^i_s}\right) ds +  \frac 12 \gamma \int_0^T  q(s)' \Sigma q(s) ds.$$

The dual problem $(\mathcal D)$ associated to this Bolza problem is:\footnote{See \cite{r1} for more details on Bolza problems.}

$$\inf_{(p^1,\ldots, p^d) \in W^{1,1}((0,T),\mathbb{R}^d) } J(p^1,\ldots,p^d),$$ where $$J(p^1,\ldots,p^d) = \sum_{i=1}^d \int_0^T V^i_s H^i\left(p^i(s)\right) ds +  \frac 1{2 \gamma} \int_0^T \dot{p}(s)' \Sigma^{-1} \dot{p}(s) ds + p(0)\cdot q_0.$$
\begin{rem}
By construction of the dual problem, the Hamiltonian equations that characterize the minimizers of $J$ are exactly those of the system $(S_H)$. In other words, any solution of problem $(\mathcal D)$ gives a (the) solution of problem $(\mathcal P)$, through $q = \frac 1{\gamma} \Sigma^{-1} \dot{p}$.\\
\end{rem}
If we consider the discrete counterpart of our model (see Appendix B), then the discrete counterpart $(\tilde{\mathcal P})$ of  $(\mathcal P)$ is:
$$\inf_{(q_n^i)_{0\le n \le N, 1 \le i \le d} \in \tilde{\mathcal{C}}} \tilde{I}((q_n^i)_{0\le n \le N, 1 \le i \le d}), $$
where
$$\tilde{I}((q_n^i)_{0\le n \le N, 1 \le i \le d}) =  \sum_{i=1}^d  \sum_{n=0}^{N-1} {L^i}\left(\frac{q_{n}^i - q_{n+1}^i}{V_{n+1}^i \Delta t} \right) V^i_{n+1} \Delta t + \frac{\gamma}{2} \sum_{n=0}^{N-1} q_{n+1}'\Sigma q_{n+1} \Delta t,$$ and where $$\tilde{\mathcal{C}} = \left\{ (q^1, \ldots, q^d)' \in \left(\mathbb{R}^{N+1}\right)^d, \forall i, q^i_0 =  q^i_0, q^i_N = 0, |q^i_n-q^i_{n+1}| \le \rho_m^i V^i_{n+1}\Delta t, 0 \le n \le N-1\right\}.$$

The dual problem $(\tilde{\mathcal D})$ associated to $(\tilde{\mathcal P})$ is:
$$\inf_{(p_n^i)_{0\le n < N, 1 \le i \le d}} \tilde{J}((p_n^i)_{0\le n < N, 1 \le i \le d}), $$
where
$$\tilde{J}((p_n^i)_{0\le n < N, 1 \le i \le d}) =  \sum_{i=1}^d  \sum_{n=0}^{N-1} V^i_{n+1} {H^i}(p_n^i) \Delta t + \frac{1}{2\gamma \Delta t} \sum_{n=1}^{N-1} (p_n - p_{n-1})\Sigma^{-1} (p_n - p_{n-1}) +  \sum_{i=1}^d p_0^i q_0^i.$$

\begin{rem}
As above, by construction of the dual problem, any solution of problem $(\tilde{\mathcal D})$ gives a solution of $(\tilde{S}_H)$ and therefore the solution of problem $(\tilde{\mathcal P})$ -- see Appendix B. This solution can be written using the relations between the primal and the dual variables: $$q_{n+1} = \frac 1{\gamma\Delta t} \Sigma^{-1}(p_{n+1}-p_n),\qquad 0\le n < N-1.$$
\end{rem}

This dual problem is at the core of our numerical approximation method since our method is based on a numerical approximation of a minimizer of $\tilde{J}$.\\

The initial idea to find such a minimizer is in fact to use a gradient descent algorithm on $\tilde{J}$. However, a simple gradient descent would be equivalent to an explicit scheme to approximate a solution of the PDE\footnote{We have indeed $\nabla J(p) = -\frac 1{\gamma} \Sigma^{-1} \ddot{p} + \left(
    \begin{array}{c}
      V^1 {H^1}'\left(p^1\right) \\
      \vdots \\
     V^d {H^d}'\left(p^d\right)
    \end{array}
  \right)$, where the gradient is taken in $L^2$.}

$$\partial_\theta p  -\frac 1{\gamma} \Sigma^{-1} \partial^2_{tt} p + \left(
    \begin{array}{c}
      V^1 {H^1}'\left(p^1\right) \\
      \vdots \\
     V^d {H^d}'\left(p^d\right)
    \end{array}
  \right)=0$$
with Neumann boundary conditions $\dot{p}(\theta,0) = \gamma \Sigma q_0$ and $\dot{p}(\theta,T) = 0$, and with an initial condition at $\theta = 0$.\\

Therefore, the na\"ive gradient descent would require a very small step for $\theta$ in order to converge to a minimum. The methodology we propose consists instead of considering a semi-implicit gradient descent.

\subsection{Semi-implicit gradient descent}

The method we propose is inspired from a gradient descent. However, we consider an implicit scheme for what would correspond to the heat operator in continuous time.\\

The idea is in fact to decompose $\tilde{J}$ as $\tilde{J}_1 + \tilde{J}_2$ where
 $$\tilde{J}_1((p_n^i)_{0\le n < N, 1 \le i \le d}) =  \frac{1}{2\gamma \Delta t} \sum_{n=1}^{N-1} (p_n - p_{n-1})\Sigma^{-1} (p_n - p_{n-1})$$
  and $$\tilde{J}_2((p_n^i)_{0\le n < N, 1 \le i \le d}) = \sum_{i=1}^d  \sum_{n=0}^{N-1} V^i_{n+1} {H^i}(p_n^i) \Delta t +  \sum_{i=1}^d p_0^i q_0^i.$$

Then, starting from an initial guess $p^0 = \left(p_0^{1,0}, \ldots, p_0^{d,0}, \ldots, p_{N-1}^{1,0}, \ldots, p_{N-1}^{d,0}\right)'$, we compute $p^{k+1}$ from $p^k$ by:

$$p^{k+1} = p^k - \frac{\Delta \theta}{\Delta t} \left(\nabla \tilde{J}_1(p^{k+1}) + \nabla \tilde{J}_2(p^k)  \right).$$

where $\Delta \theta > 0$ is to be chosen to guarantee convergence of the sequence $(p^k)_k$.\\

In other words, we propose the following semi-implicit gradient descent:

\begin{itemize}
  \item Start with an initial guess $(p_n^{i,0})_{0\le n < N, 1 \le i \le d}$
  \item For $k \ge 0$ and $\Delta \theta > 0$, define recursively\footnote{$p_n^{\cdot,k}$ denotes the $\R^d$ column vector $(p_n^{1,k}, \ldots, p_n^{d,k})'$.}  $(p_n^{i,k+1})_{0\le n < N, 1 \le i \le d}$ from $(p_n^{i,k})_{0\le n < N, 1 \le i \le d}$ by:
  $$\frac{p_n^{\cdot,k+1} - p_n^{\cdot,k}}{\Delta\theta}  - \frac 1{\gamma} \Sigma^{-1} \frac{p_{n+1}^{\cdot,k+1} - 2 p_{n}^{\cdot,k+1} + p_{n-1}^{\cdot,k+1}}{\Delta t^2} + \left(
    \begin{array}{c}
      V^1 {H^1}'\left(p_n^{1,k}\right) \\
      \vdots \\
     V^d {H^d}'\left(p_n^{d,k}\right)
    \end{array}
  \right)=0, \quad 0\le n < N,$$ where, by convention, we define:
  $$p_{-1}^{\cdot,k} = p_{0}^{\cdot,k} - \Delta t \gamma \Sigma q_0   ,\qquad p_N^{\cdot,k} = p_{N-1}^{\cdot,k}.$$
\end{itemize}

As the method is semi-implicit, we need first to state that there is no issue with the recursive definition. For that purpose, we start with a straightforward Lemma:

\begin{Lemma}
\label{j1}
For $p = \left(p_0^{1}, \ldots, p_0^{d}, \ldots, p_{N-1}^{1}, \ldots, p_{N-1}^{d}\right)'$,
$$\nabla \tilde{J}_1(p) = \frac {1}{\gamma \Delta t} M \otimes \Sigma^{-1} p$$ where the $N\times N$ matrix $M$ is defined by:

$$M =  \left( \begin{matrix}
1 & -1 & & & 0\\
-1 & 2 & -1 & &  \\
 & \ddots & \ddots & \ddots & \\
 & & -1& 2 & -1\\
0 & & & -1 & 1 \end{matrix} \right).$$
\end{Lemma}

Then, the next Proposition states that our method is indeed well defined.

\begin{Proposition}
\label{wd}
The sequence $(p^k)_k$ is well defined.
\end{Proposition}

Now, our goal is to prove that for sufficiently small values of $\Delta \theta$, the sequence $(p^k)_k$ converges toward a minimizer of $\tilde{J}$. Then, the optimal trajectory $q^*$ will also be obtained as a limit using the relation between the dual variable $p$ and the primal variable $q$. For that purpose, we start with a straightforward Lemma.

\begin{Lemma}
Assume that the Hamiltonian functions $H^i$s are $C^{1,1}$, then $\nabla \tilde{J}_2$ is a Lipschitz function with

$$\|\nabla \tilde{J}_2\|_{\mathrm{Lip}} \le \Delta t \sup_{i,n} V^{i}_{n+1} \|{H^i}'\|_{\mathrm{Lip}}.$$
\end{Lemma}

We can now state our convergence result.

\begin{Theorem}[Convergence of the semi-implicit gradient descent]
\label{convergence}
Assume that the Hamiltonian functions $H^i$s are $C^{1,1}$, and let us consider $K$ such that $\|\nabla \tilde{J}_2\|_{\mathrm{Lip}} \le K \Delta t$.\\
Then, if $\Delta \theta < \frac 2K$, $(p^k)_k$ converges towards a minimum $p^*$ of $\tilde{J}$.

Therefore, if we define
$$\forall n \in \lbrace 1, \ldots ,N-1 \rbrace, \quad q^*_n = \left(q_n^{1*}, \ldots, q_n^{d*}\right)' = \lim_{k\to +\infty} \frac{1}{\gamma \Delta t} I_{N}\otimes\Sigma^{-1}\left(p^{\cdot,k}_{n} - p^{\cdot,k}_{n-1}\right),$$ then $(q^*,p^*)$ is a solution of $(\tilde{S}_H)$.

\end{Theorem}

This theorem proves that our method works better than a simple gradient descent, as $\Delta \theta$ can be chosen independently of $\Delta t$. In addition to that, it does not require to consider second derivatives of Hamiltonian functions contrary to a Newton's scheme. Before turning to the practical use of our method, let us notice that the bound $\frac 2 K$ can be made explicit in the case of execution costs of the form $L(\rho) = \eta |\rho|^{1+\phi} + \psi |\rho|$:
\begin{rem}
If $\forall i, L^i(\rho) = \eta^i |\rho|^{1+\phi^i} + \psi^i |\rho|$, then we can consider
$$K = \sup_{i,n} V^{i}_{n+1} \frac{1}{\eta^i \phi^i(1+\phi^i)} {\rho^i_m}^{1-\phi^i}$$
\end{rem}

In practice, convergence may occur for values of $\Delta \theta$ above $\frac 2 K$.\\

\begin{rem}
If we define $q_n^k= \frac{1}{\gamma \Delta t} I_{N}\otimes\Sigma^{-1}\left(p^{\cdot,k}_{n} - p^{\cdot,k}_{n-1}\right)$, then $q_0^k = q_0$ and $q_N^k = 0$ for all $k$, thanks to the definition of $p_{-1}^k$ and $p_N^k$. This is important in practice as we shall approximate $q^*$ by  $q^k$ for a large $k$.
\end{rem}

\section{Practical examples}

\subsection{Preliminary remarks}
In practice, it may be convenient to diagonalize $\Sigma$ to simplify computations in the semi-implicit gradient descent. If we write indeed $\Sigma =  Q D Q^{-1}$ the spectral decomposition of $\Sigma$, the variable $y^k$ defined by  $y^{\cdot,k}_n = Q^{-1} p_{n}^{\cdot,k}$ verifies:

  $$\frac{y_n^{\cdot,k+1} - y_n^{\cdot,k}}{\Delta\theta}  - \frac 1{\gamma} D^{-1} \frac{y_{n+1}^{\cdot,k+1} - 2 y_{n}^{\cdot,k+1} + y_{n-1}^{\cdot,k+1}}{\Delta t^2} + Q^{-1}\left(
    \begin{array}{c}
      V^1 {H^1}'\left((Qy_n^{\cdot,k})^1\right) \\
      \vdots \\
     V^d {H^d}'\left((Qy_n^{\cdot,k})^d\right)
    \end{array}
  \right)=0, \quad 0\le n < N,$$ where, by convention, we define:
  $$y_{-1}^{\cdot,k} = y_{0}^{\cdot,k} - \Delta t \gamma DQ^{-1} q_0   ,\qquad y_N^{\cdot,k} = y_{N-1}^{\cdot,k}.$$

This formulation enables indeed to consider the discrete heat operator on each stock independently.\\

We now turn to the practical use of our method with specific examples. We shall not proceed to comparative statics as the role of the parameters have been described in many papers (see for instance \cite{almgren2003optimal,gueantframework}). Instead we focus on specific cases where the constraint on the participation rate is binding, or where bid-ask spread plays a role.\\

\subsection{Examples in the one-asset case}

To examplify the use of our method, we consider first the liquidation of a portfolio with a single stock. The parameters of the stock are the following:\footnote{The figures are inspired from the French stock Sanofi.}

\begin{itemize}
\item Stock price $S_0 = 75$,
\item Volatility 20\%, corresponding to  $\sigma =0.9375 $,
\item Market volume is assumed to be constant over the day with $V_t=2000000$,
\item The execution cost function is $L(\rho)  = \eta |\rho|^{1+\phi} + \psi |\rho|$, with $\eta = 0.045$, $\phi=0.5$ and $\psi = 0.0081$.
\end{itemize}

\begin{figure}[H]
\centering
\includegraphics[width = 0.8\textwidth]{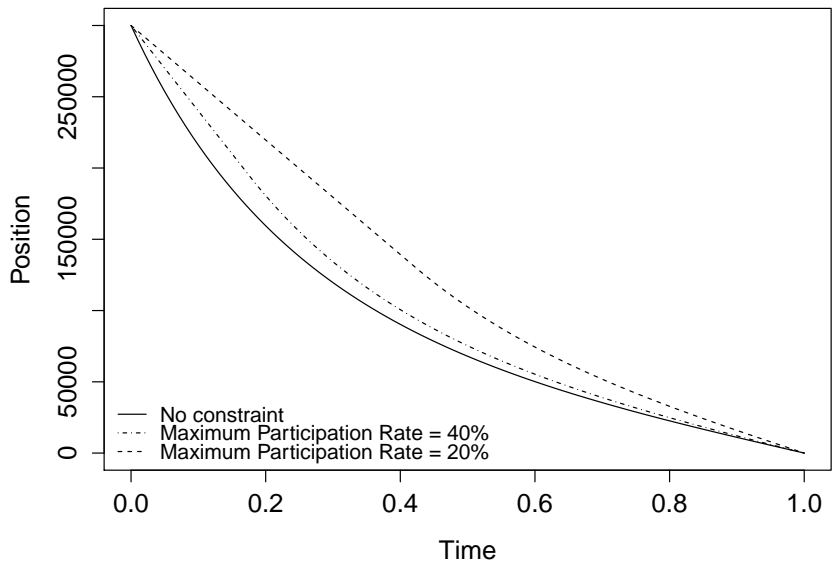}
\caption{Liquidation with different values of the maximum participation rate $\rho_m$. Risk aversion: $\gamma = 4.10^{-7}.$}
\label{1d}
\end{figure}

We consider the liquidation of $q_0 = 300000$ shares (that is $15\%$ of market daily volume) over one day $(T=1)$, with three different figures for the maximum participation rate. In the first case, we set $\rho_m = 60\%$ and since the constraint is never binding, this corresponds to setting no constraint at all on participation rate. The two other cases correspond respectively to $\rho_m = 40\%$ and $\rho_m = 20\%$. The results for the optimal liquidation strategy are shown on Figure \ref{1d}. We see that, as expected, the more we constrain the participation rate, the slower the execution process. In particular, we see that the constraint is binding in the two cases  $\rho_m = 40\%$ and $\rho_m = 20\%$, as the liquidation process starts in straight line (the slope of the line corresponding to the maximum participation rate).

\subsection{Examples in the multi-asset case}

The above examples were in the case of a one-asset portfolio. We now turn to three different cases involving several assets. In the first case, we liquidate a long portfolio of 2 positively correlated assets. In the second case, we liquidate a portfolio with a long position in the first asset and a short position in the second asset, the two assets being still positively correlated. The third case corresponds to the liquidation of a one-asset portfolio when a round trip on another (correlated) asset is allowed, in order to hedge risk.\\

We consider that the first asset is as above:

\begin{itemize}
\item Stock price $S^1_0 = 75$,
\item Volatility 20\%, corresponding to  $\sigma^1 =0.94 $,
\item Market volume is assumed to be constant over the day with $V^1_t=2000000$,
\item The execution cost function is $L^1(\rho)  = \eta^1 |\rho|^{1+\phi^1} + \psi^1 |\rho|$, with $\eta^1 = 0.045$, $\phi^1=0.5$ and $\psi^1 = 0.0081$.
\end{itemize}

For the first two cases we illustrate, the second asset we consider has the following characteristics:\footnote{The figures are inspired from the French stock Total.}

\begin{itemize}
\item Stock price $S^2_0 = 50$,
\item Volatility 17\%, corresponding to  $\sigma^2 =0.53$,
\item Market volume is assumed to be constant over the day with $V^2_t=4500000$,
\item The execution cost function is $L^2(\rho)  = \eta^2 |\rho|^{1+\phi^2} + \psi^2 |\rho|$, with $\eta^2 = 0.0255$, $\phi^2=0.5$ and $\psi^2 = 0.005$.
\end{itemize}

Correlation between the two assets is assumed to be $0.5$.\\

For the first example, we assume that $q_0^1 = 300000$ and $q_0^2 = 675000$, that is $15\%$ of the daily market volume of each asset. Maximum participation rates are assumed to be $\rho_m^1 = \rho_m^2 = 40\%$.\\

\begin{figure}[h]
\centering
\includegraphics[width = 0.8\textwidth]{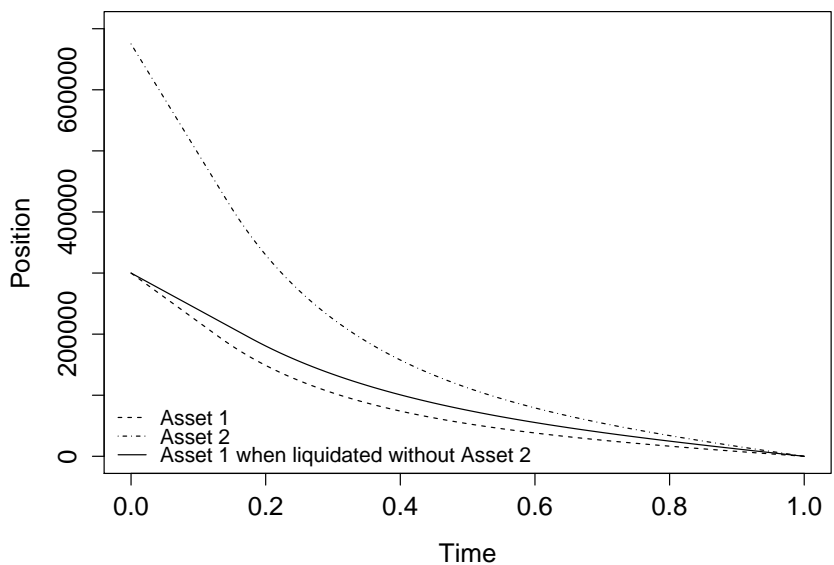}
\caption{Liquidation of a 2-asset long portfolio. Risk aversion: $\gamma = 4.10^{-7}.$}
\label{2d1}
\end{figure}

The results are shown on Figure \ref{2d1}. As a benchmark, we plotted the optimal strategy for the liquidation of the portfolio with Asset 1 only. We see that the presence of the two assets in the portfolio accelerates the liquidation of Asset 1. Since the two assets are positively correlated, price risk is increased by the presence of Asset 2. Therefore, it is natural that the liquidation of Asset 1 occurs faster in the presence of Asset 2. It is also interesting to notice that the velocity at which liquidation occurs when the full portfolio is considered would be even more important without participation contraints. The constraint on Asset 1 is indeed binding over the first 10\% of the time window in the two-asset case.\\

The opposite case where liquidation is slower in the multi-asset case than in the one-asset case may correspond either to the liquidation of a long/short portfolio with two positively correlated assets or to the liquidation of a long-only or short-only portfolio with two negatively correlated assets.\\

\begin{figure}[h]
\centering
\includegraphics[width = 0.8\textwidth]{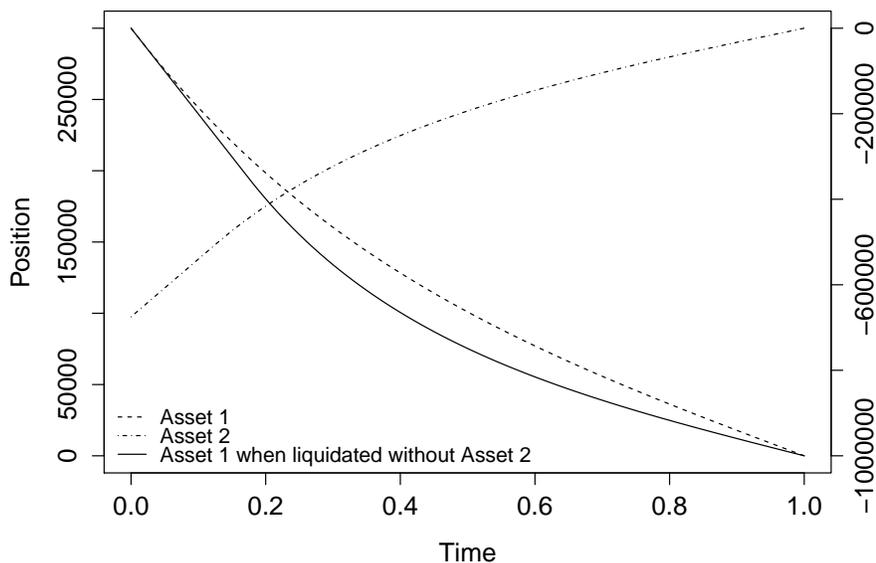}
\caption{Liquidation of a 2-asset long/short portfolio. Risk aversion: $\gamma = 4.10^{-7}.$ The left axis corresponds to Asset 1, the right axis corresponds to Asset 2.}
\label{2d2}
\end{figure}

To illustrate the former, we consider the same assets as above but the portfolio is $q_0^1 = 300000$ and $q_0^2 = -675000$. Maximum participation rates are assumed to be $\rho_m^1 = \rho_m^2 = 30\%$.\\

The results are  shown on Figure \ref{2d2}. As expected, because price risk is reduced by the presence of the second asset, the liquidation of Asset 1 is slower in the two-asset case than in the one-asset case. It is however noteworthy that the constraint on Asset 1 is binding at the very begining in both cases.\\

The third case we consider is a pure hedging case, the trader does not have an initial position on the Asset 2, but this second asset will be used for hedging purpose. We consider the case where Asset 2 is much more liquid than Asset 1:
\begin{itemize}
\item Stock price $S^2_0 = 50$,
\item Volatility 17\%, corresponding to  $\sigma^2 =0.53$,
\item Market volume is assumed to be constant over the day with $V^2_t=10000000$,
\item The execution cost function is $L^2(\rho)  = \eta^2 |\rho|^{1+\phi^2} + \psi^2 |\rho|$, with $\eta^2 = 0.002$, $\phi^2=0.5$ and $\psi^2 = 0.001$.
\end{itemize}

\begin{figure}[H]
\centering
\includegraphics[width = 0.8\textwidth]{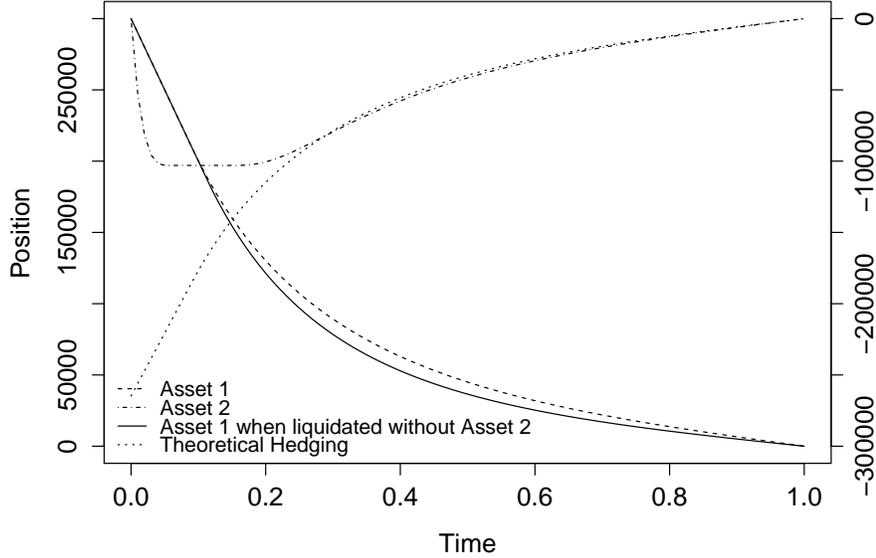}
\caption{Liquidation of a one-asset portfolio with and without hedge. Risk aversion: $\gamma = 1.10^{-6}.$ The left axis corresponds to Asset 1, the right axis corresponds to Asset 2.}
\label{2d3}
\end{figure}

Correlation between the two asset prices is assumed to be $0.5$. As far as maximum participation rates are concerned, we took $\rho^1_m = 50\%$ and $\rho^2_m$ large enough for the constraint not to be binding.\\

The results are shown on Figure \ref{2d3}. As a benchmark, we also plotted in dotted line the theoretical hedging curve $q^2=-\rho\dfrac{\sigma^1S_0^1}{\sigma^2S_0^2} q^1$, had there be no execution costs. In fact, in order to avoid paying too much for the round trip, the trader restricts its round trip by staying on a plateau (this is link to the proportional term). We also see, as expected, that the execution process is slowed down thanks to the hedging instrument.

\section*{Appendix A: Proofs}

\emph{Proof of Proposition \ref{gaussian}}:\\

After integrating by parts in the definition of $X_T$, we obtain:

$$X_T = q_0'S_0  - \int_0^T V^i_s L^i\left(\frac{v^i_s}{V^i_s}\right) ds  + \int_0^T \sigma^i q^i_s dW^i_s.$$

Hence, if $(v^1, \ldots, v^d) \in \mathcal{A}_{\textrm{det}}$, $(q^1, \ldots, q^d)$ is deterministic and $X_T$ is Gaussian. Therefore $J(v^1,\ldots, v^d) = \mathbb{E}\left[-\exp(-\gamma X_T)\right]$  can be computed in closed form as we know the Laplace transform of a Gaussian variable:

$$J(v^1, \ldots, v^d) = -\exp\left(-\gamma\left(q_0' S_0 - \sum_{i=1}^d \int_0^T V^i_s L^i\left(\frac{v^i_s}{V^i_s}\right) ds -  \frac 12 \gamma \int_0^T  q_s' \Sigma q_s ds \right)\right).$$\qed\\

\emph{Proof of Theorem \ref{main}}:\\

Existence can be obtained using the same method as in Theorem 2.1 of \cite{gueantframework}.\footnote{The proof is even simplified by the presence of participation constraints that enables to avoid the use of Dunford-Pettis Theorem.}\\

Uniqueness comes straightforwardly from the fact that the quadratic form $x \mapsto x'\Sigma x$ is a strictly convex function.\\

Now, coming to the Hamiltonian characterization, we consider the generalized functions:

$$\mathcal{L}^i(\rho) = \begin{cases} L^i(\rho)  &\mbox{if } |\rho| \le \rho^i_m \\
+\infty & \mbox{if }  |\rho| > \rho^i_m
\end{cases}, \forall i$$

Then, the problem $(\mathcal P)$ is equivalent to

$$\inf_{(q^1,\ldots, q^d) \in \widehat{\mathcal{C}}} \sum_{i=1}^d \int_0^T V^i_s \mathcal{L}^i\left(\frac{\dot{q}^i(s)}{V^i_s}\right) ds +  \frac 12 \gamma \int_0^T  q(s)' \Sigma q(s) ds$$
where $$\widehat{\mathcal{C}} = \left\lbrace q \in W^{1,1}((0,T),\mathbb{R}^d), q(0) = q_0, q(T) = 0 \right\rbrace.$$

Applying Theorem 6 of \cite{r1} and its corollary to this problem, we obtain the Hamiltonian characterization stated in Theorem \ref{main}.\qed\\

\emph{Proof of Theorem \ref{optisdet}}:\\

The proof is exactly the same as in \cite{gueantframework} and \cite{schied2010optimal}. It is a simple consequence of Girsanov Theorem.\qed\\

\emph{Proof of Proposition \ref{wd}}:\\

Using Lemma \ref{j1}, we have:

$$p^{k+1} - p^k  = -\frac {\Delta \theta}{\gamma \Delta t^2} M \otimes \Sigma^{-1} p^{k+1} - \frac{\Delta \theta}{\Delta t} \nabla \tilde{J}_2(p^k).$$

Hence, to prove that $p^{k+1}$ is uniquely defined from $p^k$, we need to prove that the matrix $$I_{Nd} + \frac {\Delta \theta}{\gamma \Delta t^2} M \otimes \Sigma^{-1}$$ is invertible. For that purpose, let us write $\Sigma = Q D Q^{-1}$ the spectral decomposition of $\Sigma$, $D$ being diagonal with positive coefficients. Notice then that

$$I_{Nd} + \frac {\Delta \theta}{\gamma \Delta t^2}M \otimes D^{-1}$$
is a strictly diagonally dominant matrix and hence it is invertible. Therefore,

$$I_{Nd} + \frac {\Delta \theta}{\gamma \Delta t^2} M \otimes \Sigma^{-1} = (I_N \otimes Q) \left(I_{Nd} + \frac {\Delta \theta}{\gamma \Delta t^2} M \otimes D^{-1}\right)(I_N \otimes Q^{-1})$$
is invertible.\qed\\

\emph{Proof of Theorem \ref{convergence}}:\\

To improve readability, let us introduce $$A = \frac 1\gamma M \otimes \Sigma^{-1}\quad  \textrm{and} \quad B = I_{Nd} + \frac {\Delta \theta}{\Delta t^2} A. $$ Then, using Lemma \ref{j1}, straightforward computations give:
$$p^{k+1} - p^k =  -\frac {\Delta \theta}{ \Delta t} B^{-1} \nabla \tilde{J}(p^k).$$

Now, we decompose $\tilde{J}(p^{k+1}) -  \tilde{J}(p^{k})$ as $\tilde{J}_1(p^{k+1}) -  \tilde{J}_1(p^{k}) + \tilde{J}_2(p^{k+1}) -  \tilde{J}_2(p^{k})$.\\

\begin{itemize}
\item For the first part, we have
$$ \tilde{J}_1(p^{k+1}) -  \tilde{J}_1(p^{k}) = \nabla\tilde{J}_1(p^{k})'(p^{k+1}-p^k) + \frac 1{2} (p^{k+1}-p^k)'\frac{A}{\Delta t}(p^{k+1}-p^k)$$
$$ =  -\frac {\Delta \theta}{ \Delta t} \nabla \tilde{J}_1(p^k)'B^{-1} \nabla \tilde{J}(p^k) +  \frac 12\left(\frac {\Delta \theta}{ \Delta t}\right)^2 \nabla \tilde{J}(p^k)'B^{-1}\frac{A}{\Delta t}B^{-1}  \nabla \tilde{J}(p^k).$$
\item For the second part, we have
$$ \tilde{J}_2(p^{k+1}) -  \tilde{J}_2(p^{k}) \le \nabla\tilde{J}_2(p^{k})'(p^{k+1}-p^k) +  \frac 1{2} K\Delta t \|p^{k+1}-p^k\|_2^2$$
$$\le  -\frac {\Delta \theta}{ \Delta t} \nabla \tilde{J}_2(p^k)'B^{-1} \nabla \tilde{J}(p^k) +  \frac 12 K\Delta t\left(\frac {\Delta \theta}{ \Delta t}\right)^2 \nabla \tilde{J}(p^k)'B^{-2}\nabla \tilde{J}(p^k).$$
\end{itemize}

Summing, we get:
$$ \tilde{J}(p^{k+1}) -  \tilde{J}(p^{k}) \le -\frac {\Delta \theta}{ \Delta t} \nabla \tilde{J}(p^k)'B^{-1} \nabla \tilde{J}(p^k) +  \frac 12\left(\frac {\Delta \theta}{ \Delta t}\right)^2 \nabla \tilde{J}(p^k)'B^{-1}\left(K\Delta t I_{Nd} + \frac{A}{\Delta t}\right)B^{-1}\nabla \tilde{J}(p^k)$$
$$\le -\frac {\Delta \theta}{ \Delta t} \nabla \tilde{J}(p^k)'B^{-1}\underbrace{\left( B -  \frac 12 \frac {\Delta \theta}{ \Delta t} \left(K\Delta t I_{Nd} + \frac{A}{\Delta t}\right)\right)}_{=R} B^{-1} \nabla \tilde{J}(p^k).$$

Hence, writing

$$\frac {\Delta \theta}{ \Delta t} \nabla \tilde{J}(p^k)'B^{-1}R B^{-1} \nabla \tilde{J}(p^k) \le    \tilde{J}(p^{k}) - \tilde{J}(p^{k+1})$$

and summing over $k$, we obtain for $\kappa \in \N$:

$$\frac {\Delta \theta}{ \Delta t} \sum_{k=0}^{\kappa}\nabla \tilde{J}(p^k)'B^{-1}R B^{-1} \nabla \tilde{J}(p^k) \le  \tilde{J}(p^{0}) - \inf  \tilde{J}.$$

Now,

$$ R = B -  \frac 12 \frac {\Delta \theta}{ \Delta t} \left(K\Delta t I_{Nd} + \frac{A}{\Delta t}\right) =  \left(1-\frac 12 K\Delta \theta\right)I_{Nd} + \frac 12 \frac {\Delta \theta}{\Delta t^2} A$$
is a positive-definite matrix for $\Delta \theta < \frac 2K$.

Therefore, the series of positive terms  $\sum_{k}\nabla \tilde{J}(p^k)'B^{-1}R B^{-1} \nabla \tilde{J}(p^k)$ is convergent.\\

As $R$ and $B$  are positive-definite matrices, we can conclude that the series $\sum_{k} \|\nabla \tilde{J}(p^k)\|_2^2$ is also convergent.\\

Now, since $p^{k+1} - p^k =  -\frac {\Delta \theta}{ \Delta t} B^{-1} \nabla \tilde{J}(p^k)$, the series $\sum_{k} \|p^{k+1}- p^k\|_2^2$ is convergent and we can conclude that the sequence $(p^k)_k$ converges toward a vector $p^*$ such that $\nabla \tilde{J}(p^*) = 0$ -- this corresponds to a minimizer of $\tilde{J}$.\\

Now, if we define $q^*_0 = q_0$, $q^*_N = 0$, and $\forall n \in \lbrace 1, \ldots ,N-1 \rbrace$,
$$q^*_n = \left(q_n^{1*}, \ldots, q_n^{d*}\right)' = \frac{1}{\gamma \Delta t} I_{N}\otimes\Sigma^{-1}\left(p^{\cdot,*}_{n} - p^{\cdot,*}_{n-1}\right),$$ it is straightforward to verify that $(p^*,q^*)$ is a solution of $(\tilde{S}_H)$.

\section*{Appendix B: Discrete model}

This appendix is dedicated to the discrete counterpart of our model. We consider for that purpose that time is divided into slices of length $\Delta t$ and we denote by $t_0 = 0 \le\ldots \le  t_n = n\Delta t \le \ldots \le t_N = T $ the relevant sequence of times for our discrete model.\\ 

For $i \in \{1, \ldots, d\}$, we denote by $v^i_{n+1} \Delta t$ the number of shares\footnote{The fact that deterministic strategies are optimal can be shown in the discrete framework using similar techniques as in the continuous-time model.} of stock $i$ sold by the trader between $t_{n}$ and $t_{n+1}$. As a consequence, the state of the portfolio $q = (q^1, \ldots, q^d)$ is given by

$$\forall i, q^i_{n+1} = q^i_n - v^i_{n+1} \Delta t, \quad 0\le n < N.$$

Price processes are modeled by:

$$S^i_{n+1} = S^i_n + \sigma^i \sqrt{\Delta t}  \epsilon^i_{n+1},$$

where $(\sigma^1 \epsilon^1_n, \ldots, \sigma^d \epsilon^d_n)_{n}$ are \emph{i.i.d.} $\mathcal{N}(0,\Sigma)$ random variables.\\

The amount of cash obtained by the trader for the $v^i_{n+1}\Delta t$ shares of stock $i$ he sold over $(t_n,t_{n+1}]$ depends on $v^i_{n+1}\Delta t$ itself and on the market volume for stock $i$ over $(t_n,t_{n+1}]$, assumed to be $V^i_{n+1}\Delta t$.\\

The resulting dynamics for the cash account is:

$$X_{n+1} = X_n + \sum_{i=1}^d \left(v^i_{n+1} S^i_n - L^i\left(\frac{v^i_{n+1}}{V^i_{n+1}}\right)V^i_{n+1}{\Delta t}\right), \qquad X_0 = 0.$$

The maximization criterion we consider is:

$$\mathbb{E}\left[-\exp(-\gamma X_N)\right].$$

As in the continuous-time model, the final wealth can be computed:

$$X_N = \sum_{i=1}^d \left(q^i_0 S^i_0 + \sigma^i \sqrt{\Delta t}  \sum_{n=0}^{N-1} q^i_{n+1} \epsilon^i_{n+1} - \sum_{n=0}^{N-1} L^i\left(\frac{v^i_{n+1}}{V^i_{n+1}}\right)V^i_{n+1}{\Delta t}\right).$$

Hence, $X_N$ is a Gaussian variable with mean $$\sum_{i=1}^d \left(q^i_0 S^i_0 - \sum_{n=0}^{N-1} L^i\left(\frac{v^i_{n+1}}{V^i_{n+1}}\right)V^i_{n+1}{\Delta t}\right)$$ and variance
$${\Delta t}  \sum_{n=0}^{N-1} q_{n+1}' \Sigma q_{n+1}.$$

Therefore,
$$\mathbb{E}\left[-\exp(-\gamma X_N)\right]$$
$$= -\exp\left(-\gamma\left(\sum_{i=1}^d \left(q^i_0 S^i_0 - \sum_{n=0}^{N-1} L^i\left(\frac{v^i_{n+1}}{V^i_{n+1}}\right)V^i_{n+1}{\Delta t}\right) - \frac\gamma 2 {\Delta t}  \sum_{n=0}^{N-1} q_{n+1}' \Sigma q_{n+1} \right)\right),$$ and the problem boils down to minimizing

$$  \sum_{i=1}^{d} \sum_{n=0}^{N-1} \left(L^i\left(\frac{q^i_n-q^i_{n+1}}{V^i_{n+1}\Delta t}\right)V^i_{n+1}{\Delta t} \right)+ \frac{\gamma}{2}\sum_{n=0}^{N-1} q_{n+1}' \Sigma q_{n+1} {\Delta t},$$
over $$\tilde{\mathcal{C}} = \left\{ (q^1, \ldots, q^d)' \in \left(\mathbb{R}^{N+1}\right)^d, \forall i, q^i_0 =  q^i_0, q^i_N = 0, |q^i_n-q^i_{n+1}| \le \rho_m^i V^i_{n+1}\Delta t, 0 \le n \le N-1\right\}.$$

This problem can be written as a discrete-time Bolza problem:

$$\inf_{(q_n^i)_{0\le n \le N, 1 \le i \le d}, q_0=q_0, q_N= 0}\mathcal{I}((q_n^i)_{0\le n \le N, 1 \le i \le d}),$$
where $$\mathcal{I}((q_n^i)_{0\le n \le N, 1 \le i \le d}) =  \sum_{n=0}^{N-1} \left(\mathcal{L}^i\left(\frac{q^i_n-q^i_{n+1}}{V^i_{n+1}\Delta t}\right)V^i_{n+1}{\Delta t} \right)+ \frac{\gamma}{2}\sum_{n=0}^{N-1} q_{n+1}' \Sigma q_{n+1} {\Delta t},$$

with  $$\mathcal{L}^i(\rho) = \begin{cases} L^i(\rho)  &\mbox{if } |\rho| \le \rho^i_m \\
+\infty & \mbox{if }  |\rho| > \rho^i_m
\end{cases}.$$

Then, using the same techniques as those developed in \cite{lasry}, but in discrete time, one can show that the dual formulation of this problem is problem $(\tilde{\mathcal D})$:
$$\inf_{(p_n^i)_{0\le n < N, 1 \le i \le d}} \tilde{J}((p_n^i)_{0\le n < N, 1 \le i \le d}), $$
where
$$\tilde{J}((p_n^i)_{0\le n < N, 1 \le i \le d}) =  \sum_{i=1}^d  \sum_{n=0}^{N-1} V^i_{n+1} {H^i}(p_n^i) \Delta t + \frac{1}{2\gamma \Delta t} \sum_{n=1}^{N-1} (p_n - p_{n-1})\Sigma^{-1} (p_n - p_{n-1}) +  \sum_{i=1}^d p_0^i q_0^i.$$

The first order condition associated to these two problems correspond to the Hamiltonian equations $(\tilde{S}_H)$:\footnote{As the function $H^i$s are of class $C^1$, the derivation of this system is straightforward for the dual problem. It is also straightforward for the primal problem when the functions $L^i$s are of class $C^1$ and when there is no participation constraint. However, in the general case, one needs to rely on classical techniques of convex optimization -- such as the very general ones presented in \cite{lasry} -- and mimic the (rather long but classical) proofs in the continuous-time case to obtain the equivalence between the primal problem, the dual problem and the Hamiltonian equations, in the discrete-time case.}
\[
(\tilde{S}_H): \quad \left \{
\begin{array}{c c l}
    p_{n+1} & = & p_n + \Delta t \gamma \Sigma q_{n+1}, \quad 0\le n <N-1\\
    q^i_{n+1} & = & q^i_n  + \Delta t V^i_{n+1} {H^i}'(p^i_n) , \quad 0\le n <N, \quad \forall i \in \{ 1, \ldots, d\}\\

\end{array}
\right.\] $$ q_0 = q_0, \quad q_N = 0.$$

\bibliographystyle{plain}

\end{document}